\documentclass{sig-alternate}
\usepackage{relsize}
\usepackage{tabularx}
\usepackage{booktabs}
\usepackage{graphicx}
\usepackage{algorithm}
\usepackage{algpseudocode}
\usepackage{pifont}
\usepackage{array}

\newcolumntype{L}{>{\centering\arraybackslash}m{3cm}}

\begin{document}
%

\title{Two-stage Cascaded Classifier for Purchase Prediction
}
%
%
%
%
%

\numberofauthors{2} 
%
\author{
%
%
\alignauthor
Sheikh Muhammad Sarwar\\  
       \affaddr{University of Dhaka,  Bangladesh}\\
       \email{smsarwar@du.ac.bd (DU\_Erudite team)}
\alignauthor
Mahamudul Hasan\\ 
       \affaddr{University of Dhaka, Bangladesh}\\
       \email{munna09bd@gmail.com}
\and
\alignauthor 
Dmitry I. Ignatov\\
       \affaddr{National Research University}\\ 
			\affaddr{Higher School of Economics, Moscow, Russia}\\
       \email{dignatov@hse.ru}
		}

\maketitle
\begin{abstract}
In this paper we describe our machine learning solution for the RecSys Challenge 2015. We have proposed a time-efficient two-stage cascaded classifier for the prediction of buy sessions and purchased items within such sessions. Based on the model, several interesting features found, and formation of our own test bed, we have achieved a reasonable score. Usage of Random Forests helps us to cope with the effect of the multiplicity of good models depending on varying subsets of features in the purchased items prediction and, in its turn, boosting is used as a suitable technique to overcome severe class imbalance of the buy-session prediction.
\end{abstract}

\category{1.2.6}{Artificial Intelligence}{Learning}



\section{Introduction}
For us, the most challenging part of the competition is the classical problem in machine learning, \emph{class imbalance} \cite{Seiffert:2009}. The task of the challenge is to predict ``buy sessions'' from a large set of sessions. Each session contains the click data of several anonymous users. However, only 5\% of the training data consist of buy sessions, which is actually half a million among 9.3 million sessions. Hence, if we consider two classes (\emph{buy} and \emph{non-buy}), a severe class imbalance problem appears. Apart from predicting the buy sessions, the challenge organizers also want to know which items would be bought from a buy session. In order to answer these two questions, we have come up with two machine learning models, where one model is able to select the buy sessions and the other one finds the related items. For finding buy sessions while handling the class imbalance problem, we use the Adaboost.M1 implementation from Weka machine learning framework. We have finally found interesting features and achieved a reasonable score of 45,027 in the challenge.

\section{Problem Statement}
\label{sec:task}
RecSys Challenge 2015 provides a data set containing the clickstream data of several users of an e-commerce site over the duration of six months. The training data set is composed of a set of sessions $S$ and each session $s\in S$ contains a set of items $I_s$. $B_s\subseteq  I_s$ denotes the set of items which has been bought in session $s$. There are two types of sessions $S_b$ (the sessions that end in buying) and $S_{nb}$ (the sessions that do not end in buying) as follows:
$$S_b = \{s\in S: B_s \neq \emptyset \}, \mbox{ and } S_{nb} = S \setminus S_b.$$

The test data also contains a set of sessions $S_t$ and $S_t\cap S = \emptyset$. Now, given the set of sessions $S_t$, our task is to find all the sessions $S_{tb} \subseteq S_t$, which have at least one buy event. Moreover, if a session $s$ contains a buy event, we have to predict the set of items $B_s$ that was bought. The final solution should contain sessions $Sl$ and the set of items $A_s$ for each session $s \in Sl$, which should be bought. There is an original solution file, which contains a set of sessions $S_b$, and for each session $s \in S_b$ there is a set of items $B_s$ which were bought in that session. However, rather than maximizing any traditional evaluation function we will have to maximize a unique scoring function based on the original solution file, which is shown below:


\[
	score(S_t) = 
	‎‎\sum\limits_{\forall s \in Sl} (-1)^{[s \not\in S_b]} \frac{|S_b|}{|S_t|}+[s \in S_b] \frac{|A_s \cap B_s|}{|A_s \cup B_s|}, \mbox{ where}
\]

$[Z]$ equals 1 if $Z$ is true and 0 otherwise (Iverson notation).

\subsection {Observation about the Scoring Function}
According to the challenge description, buying event occurs in only 5\% of the sessions and that's why we can assume that the penalty for predicting wrong sessions is only 0.05. From the other hand, for predicting a right session it is possible to achieve 1.05 score and as a consequence the strategy should be to identify most of the sessions as \emph{buy sessions}. However, there is a limitation in the solution file size (<25MB) and that is why a proper level of precision should be reached in choosing the resulting sessions.  

\subsection{Data Description}
The task contains three data files: click data file, buy data file and test data file. The click data file contains all the click sessions ($S_c$) and some information about the items clicked in those sessions, while the buy data file contains a set of sessions ($S_b$) in which at least one item has been bought. The click data file contains a set of tuples in the form \{\emph{sessionId, timestamp, itemId, category}\} and the buy data file contains a set of tuples in the form \{\emph{sessionId, timestamp, itemId, price, quantity}\}. The test data file follows the same data schema as the click data file. The data description is provided in Table \ref{my-label}.

\begin{table}[!h]
\centering
\caption{Data Description}
\label{my-label}
\noindent\begin{tabularx}{\linewidth}{|c|c|X|}
\hline
File(s) type            & Attributes & \multicolumn{1}{c|}{Description}                                                                                                                                                                                  \\ \hline
click and buy & sessionId  & Unique session id                                                                                                                                                                             \\ \hline
click and buy & itemId     & Unique item identifier                                                                                                                                                                                   \\ \hline
click and buy & timestamp  & Time of click event or buy event for an item. It features day, month, hour and time of the specific event.                                                                                                           \\ \hline
click file           & category   & The category of an item in a session. 0 indicates unknown category and $1,\ldots, 12$ denote regular categories. Other numbers indicate a brand, while ``S'' is the item from a special category. \\ \hline
buy file             & price      & Price of an item (0 when the price is not available)                                                                                                                                  \\ \hline
buy file      & quantity   & Number of times a particular item has been bought in that session                                                                                                                                                             \\ \hline
\end{tabularx}
\end{table}

\section{Data Preprocessing}
\subsection{Resolving Missing Category Information}
\label{sec:category}
At first we extract several properties of items and assign 0 as category number to the items with unknown category (almost half of the sessions items belong to this category). In fact the data were collected over six months period -- from April to September, where most of the category information appeared after mid of June. We assume that the e-commerce site could not provide category data for that period of time. However, as an item repeatedly occurred in different sessions, it is possible to recover the category for most of the items using the data from June to September. While resolving the category we try to recover only regular category, i.e. we do not use  special category . However, from our analysis it was evident that an item belongs to several categories. Nonetheless, we want to find a specific category for a specific item. So, we resolve item categories using the following rule: \emph{if an item belongs to several categories, the actual category is its most frequent one in the click data}. After resolving the categories we append the attribute originalCategory to each click data.   

\subsection{Performing Temporal Ordering}
At first, we sort both the click file and buy file using sessionId, which has a specific benefit that we will discuss later. Then for each of the sessions belonging to the files we sort the session data by timestamp. By so doing we can find the temporal ordering in the click and buy data. This temporal ordering helps us determine the order of user's clicks on the items in a session. Moreover, the duration of a click could easily be found by simply subtracting time of that click from the time of the next click. Now, for each distinct item in a session if we sum the duration of the clicks in which the item appears, we define the duration of the item in that session. After sorting by timestamp we append  itemDuration (the amount of time an item is inspected in a session) to each click data.

\subsection{Extraction of Item Properties}
We extract several other properties, which are specific to an item and append it to each click data. The properties are listed in Table \ref{Item Properties}. At first we tried to solve the problem using only click-buy ratio and obtained a score of 29000. In that process, we averaged the click-buy ratio of all the items in a session and if the average was over the threshold of 5.5 we identified that session as a buy one. After that we picked half of the items by sorting on click-buy ratio. Using this simple approach we understood that we need to build features using this valuable statistic for developing our machine learning solution.

\begin{table}[h]
\centering
\caption{Item Specific Properties}
\label{Item Properties}
\noindent\begin{tabularx}{\linewidth}{|c|X|}
\hline
Item property   & Description                                                                         \\ \hline
click-buy ratio & buyCount/clickCount                                                                 \\ \hline
clickCount      & The number of buy sessions for an item. For clickCount=0  we increase it by 1 \\ \hline
buyCount        & The number of click sessions for an item. For buyCount=0 we increase it by 10       \\ \hline
price           & The price of an item. For price=0 we increase it by 1000.                      \\ \hline
itemDuration           & The total time spent on an item over all the sessions in training set                      \\ \hline
\end{tabularx}
\end{table}

\vspace*{-\baselineskip}
\begin{table}[h]
\centering
\caption{Data Properties}
\label{data_properties}
\noindent\begin{tabularx}{\linewidth}{|X|c|}
\hline
Statistics of training data                                      & Description \\ \hline
Number of buy sessions ($n_b$)                              & 509696      \\ \hline
Number of non-buy sessions ($n_c)$                           & 8740001     \\ \hline
Distinct (item, buy session) pairs & 1049817     \\ \hline
Distinct (item, non-buy session) pairs & 25565682    \\ \hline
Number of items in training data                                  & 52146       \\ \hline
\end{tabularx}
\end{table}

\subsection{Development of an Alternative Testbed}
\label{sec:testbed}
We have developed a complete framework in Java for performing the task, which includes the development of our own testbed. The development of our own independent testbed is crucial since there are only three chances of submission in the challenge per 24-hours. In order to do that we put each buy session $s$ from all the sessions in our click data $S_c$ to a file \emph{clickBuy} (click sessions ended in buy) and all non-buy sessions are placed to the file \emph{onlyClick} (click sessions not ended in buy). We sort all the click and buy sessions by sessionId in advance and split them finally (within $O(|S_c| log|S_c|)$).

After splitting, we randomly take half of the sessions in the clickBuy file to create our local test set and solution file; the remaining half of the data from clickBuy file is left for training our model and its evaluation in our own testbed. Finally, we add randomly chosen one fourth of sessions from the onlyClick file to our new test set; the remaining goes to our new training set. As a result, we have our own original testing file and test (solution) file using which we can estimate the performance of our classifier. The number of sessions in our own test data is 2439481, while the original test data contains 2312432 sessions. Moreover, we kept 254510 buy sessions in our test data and according to scoring function shown in Section \ref{sec:task}, we can achieve at most $254510 \times 1.05 = 267235.5$ score from our own test data.  

\section{Proposed Solution}
The task is intuitively divided in to two subtasks: predicting the outcome of a session and given a session predicting the set of items that should be bought in that session. So, we construct two classifiers to address these two subtasks. However, at first we thought of building a single classifier that would extract features from the items from the sessions in training data, and, given the items of a session in test data, it should classify the item as \emph{buy} or \emph{non-buy}. Unfortunately, in this process we have to handle a large feature data. If we look at Table \ref{data_properties}, it is evident that developing only an item-based classifier would require to build a model for 1049817 items labeled with buy and 25565682 items marked as non-buy. Building any sophisticated classifier around that amount of data would take a huge computation time. Hence, at first we predict the sessions, which would end in buy, and then look for the prospective bought items in those sessions. So, we need to develop two classifiers: \emph{item classifier} and \emph{session classifier}.

\subsection{Item Classifier}
An important observation about item classifier is that we have to train the classifier only with the click data of the buy sessions in training data. The click data of a buy session contain a set of items that was bought ($B_s$) and a set of items that was not bought ($A_s$). Now for each item $i \in B_s$ we extract both session-based and item-based features. After that we label that item as \emph{buy} and each item $i \in A_s$ as \emph{non-buy}. By considering only the buy sessions we get 1049817 bought items and 1264870 non-bought ones. As a result, the class imbalance problem is less stressed and the classifier has to be trained with only 215053 data in total, which results in short training time.

Since we obtained our own testbed (see Section~\ref{sec:testbed}), we have our local solution file, which we can use to verify the performance of our method.  We have tried to predict items with the classifier given the sessions in the local solution file i.e. we pretend that we know the correct sessions beforehand and we only try to test the efficiency of our item classifier. The maximum achievable score from our local solution file is 267235.5. Since we have met Rashomon effect (multiplicity of good subsets of features/models) in this item classification, we use Random Forests (RF) to cope with this difficulty \cite{Breiman:2001}; moreover, RF classifier has shown the best score in our testbed and the highest Average Jaccard measure (see Table~\ref{tbl:itemclass}). The performance of the classifiers in shown in \ref{tbl:itemclass} and the selected features are described in Table \ref{item_features}. In order to extract features 9,10,14,15,21 and 21 we cluster session data using category that we resolved previously in \ref{sec:category}.

\begin{table}[!htb]
\centering
\caption{Item Features}
\label{item_features}
\noindent\begin{tabularx}{\linewidth}{|m{0.5cm}|X|}
\hline
No. & Feature name and description \\ \hline
1 & click-buy ratio: click-buy ratio of an item\\ \hline
2 & numberOfAppearance: number of times an item appeared in a session\\ \hline
3 & itemPosition: the position of an item in the session after temporal ordering \\ \hline
4 & isSunday: true if the item's session occurred on Sunday\\ \hline
5 & isTuesday: true if the item's session occurred on Tuesday\\ \hline
6 & hour: hour of the item's session\\ \hline
7 & numberOfItems: number of distinct items in the item's session\\ \hline
8 & itemAppearanceOverThree: number of items in a session appearing more than thrice in it \\ \hline
9 & isFirstItemCategory: true if the item is the first one clicked in a specific category of the session.\\ \hline
10 & isLastItemCategory: true if the item is the last one clicked in a specific category of the session.\\ \hline
11 & buyCount: number of sessions in training data where the item is bought \\ \hline
12 & itemDuration: number of seconds spent on the item in the specific session \\ \hline
13 & price: price of the item\\ \hline
14 & categoryTopClickBuyRatio: 1 if the click-buy ratio of the item is the highest within its category and session, otherwise 0\\ \hline
15 & categoryTopBuy: 1 if the buyCount of the item is the highest within its category and session, otherwise 0\\ \hline
16 & sessionTopClickBuyRatio: 1 if the click-buy ratio of the item is the session highest, otherwise 0\\ \hline
17 & sessionTopBuyCount: 1 if the buyCount of the item is the highest in its session, otherwise 0\\ \hline
18 & maxDuration: 1 if the duration of the item is the highest in its category and session, otherwise 0\\ \hline
19 & itemOwnDuration: is the period of time that the item has been observed in the training data\\ \hline
20 & clickCount: number of sessions in training data where the item has been clicked\\ \hline
21 & categoryLowestPrice: 1 if the price of the item is the lowest within its category and session, otherwise 0\\ \hline
22 & categoryHighestPrice:  1 if the price of the item is the lowest in its category and session, otherwise 0\\ \hline
\end{tabularx}
\end{table}

\begin{table}[!htb]
\centering
\caption{Item Classifier Performance}
\label{tbl:itemclass}
\noindent\begin{tabularx}{\linewidth}{|m{2.2cm}|X|X|X|m{0.7cm}|}
\hline
Classifier & Score  & Possible score & Average Jaccard & Time to build \\ \hline
Na\"{\i}ve Bayes & 172211 & 267235.5 & 0.699 & \textbf{16 s}  \\ \hline
Bayes net & 182226 & 267235.5 &  0.727 & 65 s\\ \hline
Logistic regression & 186561 &  267235.5 &  0.751 & 123 s\\ \hline
Random forest & \textbf{189537} & 267235.5 & \textbf{0.767} & 218 s\\ \hline
\end{tabularx}
\end{table}

\subsection{Session Classifier}
We have 509696 buy sessions and 8740001 non-buy ones in the training data. Rather than looking at the feature of the items, we look at the features of sessions. There are two important observations for building the session classifier: 
 1) Since we have only 0.05 penalty for selecting a non-buy session, it should be a high recall classifier; 
 2)  As there exists class imbalance problem, a classifier would tend to predict most of the sessions as \emph{non-buy}.

In order to address the problems we performed resampling of data and classified sessions using AdaBoost.M1 algorithm. AdaboostM1 comes as a remedy for class imbalance problem combined with resampling \cite{Seiffert:2009}, and it gives us the highest recall among the classifiers we tested. Table \ref{tab:session_classifier} shows that even though RF works better when classifying items, it cannot reach high recall when classifying sessions; moreover, it takes the highest time for model training. For Adaboost.M1 the model is built in a smaller period of time (3 min 2 s for 1896886 feature vectors) and evaluation is done in 23 s for 2312432 feature vectors. We have experimented with Adaboost.M1 using different set of features; we describe the selected features in Table~\ref{tab:session_features}. In Table~\ref{tab:session_selected_features} we show the performance of Adaboost.M1 with the set of all the extracted features and it produces the highest score -- 107763.7. However, in our local testbed it returns a huge number of sessions that cannot be accommodated in the solution file of less than 25 MB (the challenge constraint). From Table~\ref{tab:session_selected_features} one can observe that with selected features we obtain a score of 106508.4, which is the highest considering the file size. Finally, we have obtained our best challenge score 45027 by cascading Adaboost.M1 with resampling for session classification and RF for choosing items.

In Table~\ref{tab:session_selected_features} we show the result of considering different subsets of features obtained from Table~\ref{tab:session_features}. We excluded the time based features, aggregated features, and other subsets to evaluate the session classifier. Moreover, we mention two of the challenge scores and show the respective score from our own testbed. As there are 2312432 session in the challenge test data, we can assume that there are 115622 (5\%) buy sessions in it and the Maximum Possible Score (MPS) from the test data can be 115622 $\times$ 1.05 $=$121402, approximately. However, we kept 254510 buy sessions in our test data and the MPS from our testbed can be 267235.5 (see Table \ref{tbl:itemclass}). As shown in Table \ref{tab:session_selected_features}, we achieve our best score 45027 (37\% of challenge test data), when our testbed score is 106508 (39\% of our own test data). When we achieve 39127 (32\% of challenge test data), our testbed score is 96077 (35\% of our own test data). So, we can assert that we are able to approximate our challenge score from our testbed one.   

\begin{table}[!htb]
\centering
\caption{Selected Session Features}
\label{tab:session_features}
\noindent\begin{tabularx}{\linewidth}{|m{0.5cm}|X|}
\hline
No. & Feature name and description\\ \hline
1 & numberOfClicks: number of clicks in a session \\ \hline
2 & numberOfItems: number of distinct items in a session \\ \hline
3 & itemsOverClickBuyRatio: number of items having click-buy ratio over 3.6, median click-buy ratio of all items\\ \hline
4 & itemsOverClickBuyCount: number of items having click-buy count over 57, average buyCount of all items\\ \hline
5 & averageClickBuyRatio: average click-buy ratio over all the items in a session \\ \hline
6 & averageBuyCount: average buy count for all the items in a session \\ \hline
7 & itemAppearanceOverTwo: number of items in a session appearing more than twice in it\\ \hline
8 & itemAppearanceOverThree: number of items in a session appearing more than three times in it \\ \hline
9 & duration: duration of a session \\ \hline
10 & hour: hour of the day when the session occurred \\ \hline
11 & isSunday: true if the session day is Sunday \\ \hline
12 & isTuesday: true if the session day is Tuesday \\ \hline
13 & averageItemDuration: itemDuration for an item is the amount of time spent on that item in the training data. This feature is the average over itemDuration of each item in a session\\ \hline
14 & averageItemClickCount: clickCount for an item is the number of times an item has been clicked in the training data. This feature is the average of clickCount of each item in a session\\ \hline
15 & averageItemPrice: average item price in a session\\ \hline
\end{tabularx}
\end{table}


\begin{table}
\centering
\setlength{\tabcolsep}{2pt}
\caption{Feature Selection for Session Classifier}
\label{tab:session_selected_features}
\noindent\begin{tabular}{|l|c|c|c|c|c|c|c|}
\hline
Selected & $R$ & $P$ & $F_1$ & Overall   & Final  &  Number of   \\
features &  &  &  &  score &  score  &  sessions  \\  \hline
all (>25MB)   & \textbf{0.77} & 0.32 & 0.35 & \textbf{107764} &  n/a & \textbf{ 963307}\\ \hline
selected  & 0.71 & 0.23 & 0.34 & 106508 &  \textbf{45027} &  782337\\ \hline
w/o time-based  & 0.60 & \textbf{0.27} & \textbf{0.37} & 88445 &  n/a &  563443 \\ \hline
w/o 3 and 4 & 0.72 & 0.22 & 0.36 & 103300 &  n/a &  832275 \\ \hline
{1,5,6,7,15} & 0.67 & 0.22 & 0.34 & 96077 &   39127 &  765873 \\ \hline
w/o  1 and 2 & 0.69 & 0.24 & 0.35 & 103944 &  n/a &  744994 \\ \hline
w/o aggregated & 0.66 & 0.25 & 0.36 & 98011 &  n/a &  684560 \\ \hline
\end{tabular}
\end{table}
\setlength{\textfloatsep}{1cm}

\begin{table}
\centering
\setlength{\tabcolsep}{2.5pt}
\caption{Session Classifier Performance}
\label{tab:session_classifier}
\noindent\begin{tabular}{|c|c|c|c|c|c|} 
\hline
Classifier & Score  & Possible  & $R$ & $P$ & Model \\
 &  &  score & & & building\\ 

\hline
Na\"{\i}ve Bayes & 49890.8 & 267235.5 & 0.36 & \textbf{0.30} & \textbf{30 s} \\ \hline
Random Forest & 87854.6 & 267235.5 & 0.58 & 0.26 & 1308 s\\ \hline
Bayes Net & 90865.5 & 267235.5 & 0.66 & 0.23 & 146 s\\ \hline
Logistic Regression & 100134.6 &  267235.5 & 0.67 & 0.25 & 152 s \\ \hline
Adaboost.M1 & \textbf{106508.3} & 267235.5 & \textbf{0.71} & 0.23 & 178 s \\ \hline 
\end{tabular}
\end{table}

\section{Conclusions}
It seems the proposed approach can be applied in a similar clickstream classification setting. The usage of a cascade of two different classifiers is beneficial  when one experiences severe imbalance between buy and non-buy sessions and multiplicity of good feature subsets. Moreover, learning and classification phases are reasonably short for this subtask decomposition. We have no doubts that further model tuning would definitely allow to achieve a better score.
%
\bibliographystyle{abbrv}
\bibliography{sig-alternate}  
%
%

\end{document}